\documentclass[
 reprint,
 superscriptaddress,
 nofootinbib,
 aps,
pra,
longbibliography
]{revtex4-2}

\usepackage{graphicx} 
\usepackage{amsmath,amssymb}
\usepackage{braket}
\usepackage[dvipsnames]{xcolor}
\usepackage[
  colorlinks=true,
linkcolor=Blue,
  citecolor = Blue,
  urlcolor = Blue,
  unicode
  ]{hyperref}
\usepackage{natbib}
\usepackage[nameinlink]{cleveref}
\crefname{equation}{Eq.}{Eqs.}
\Crefname{equation}{Equation}{Equations}
\crefname{figure}{Fig.}{Figs.}

\Crefname{figure}{Figure}{Figures}

\usepackage{ulem}

\begin{document}

\preprint{APS/123-QED}

\title{Approaching the Limit in Multiparameter AC Magnetometry with Quantum Control}

\author{Takuya Isogawa} 
\thanks{These authors contributed equally.}
\affiliation{Research Laboratory of Electronics, Massachusetts Institute of Technology, Cambridge, MA 02139, USA}
\affiliation{Department of Nuclear Science and Engineering, Massachusetts Institute of Technology, Cambridge, MA 02139, USA}

\author{Zhiyao Hu} 
\thanks{These authors contributed equally.}
\affiliation{Pritzker School of Molecular Engineering, The University of Chicago, Chicago, IL 60637, USA}

\author{Ayumi Kanamoto} 

\thanks{These authors contributed equally.}
\affiliation{Research Laboratory of Electronics, Massachusetts Institute of Technology, Cambridge, MA 02139, USA}
\affiliation{Department of Nuclear Science and Engineering, Massachusetts Institute of Technology, Cambridge, MA 02139, USA}
\affiliation{Department of Electrical and Electronic Engineering, Institute of Science Tokyo, Meguro,
Tokyo 152-8550, Japan}

\author{Nutdech Phadetsuwannukun
}
\affiliation{Research Laboratory of Electronics, Massachusetts Institute of Technology, Cambridge, MA 02139, USA}
\affiliation{Department of Physics, Massachusetts Institute of Technology, Cambridge, MA 02139, USA}

\author{Shilin Wang}
\affiliation{
   Department of Mechanical and Automation Engineering, The Chinese University of Hong Kong, Shatin, Hong Kong}

\author{Shunsuke Nishimura} 
\affiliation{Research Laboratory of Electronics, Massachusetts Institute of Technology, Cambridge, MA 02139, USA}
\affiliation{Department of Nuclear Science and Engineering, Massachusetts Institute of Technology, Cambridge, MA 02139, USA}
\affiliation{Department of Physics, The University of Tokyo, Bunkyo-ku, Tokyo, 113-0033, Japan}

\author{Boning Li} 
\affiliation{Research Laboratory of Electronics, Massachusetts Institute of Technology, Cambridge, MA 02139, USA}
\affiliation{Department of Physics, Massachusetts Institute of Technology, Cambridge, MA 02139, USA}

\author{Liang Jiang} 
\affiliation{Pritzker School of Molecular Engineering, The University of Chicago, Chicago, IL 60637, USA}

\author{Zain H. Saleem} 
\affiliation{Mathematics and Computer Science Division, Argonne National Laboratory, Lemont, IL, USA}

\author{Guoqing Wang} 
\affiliation{Research Laboratory of Electronics, Massachusetts Institute of Technology, Cambridge, MA 02139, USA}
\affiliation{Department of Physics, Massachusetts Institute of Technology, Cambridge, MA 02139, USA}

\author{Haidong Yuan} 
\email{hdyuan@mae.cuhk.edu.hk}
\affiliation{
   Department of Mechanical and Automation Engineering, The Chinese University of Hong Kong, Shatin, Hong Kong}
\affiliation{The Hong Kong Institute of Quantum Information Science and Technology, The Chinese University of Hong Kong, Shatin, Hong Kong SAR, China}
\affiliation{State Key Laboratory of Quantum Information Technologies and Materials, The Chinese University of Hong Kong, Shatin, Hong Kong SAR, China}   
\author{Paola Cappellaro} 
\email{pcappell@mit.edu}
\affiliation{Research Laboratory of Electronics, Massachusetts Institute of Technology, Cambridge, MA 02139, USA}
\affiliation{Department of Nuclear Science and Engineering, Massachusetts Institute of Technology, Cambridge, MA 02139, USA}
\affiliation{Department of Physics, Massachusetts Institute of Technology, Cambridge, MA 02139, USA}

\date{\today}

\begin{abstract}
Simultaneously estimating multiple parameters at the ultimate limit is a central challenge in quantum metrology, often hindered by inherent incompatibilities in optimal estimation strategies. At its most extreme, this incompatibility culminates in a fundamental impossibility when the quantum Fisher information matrix (QFIM) becomes singular, rendering joint estimation unattainable. This is the case for a canonical problem: estimating the amplitude and frequency of an AC magnetic field, where the generators are parallel to each other. Here, we introduce a quantum control protocol that resolves this singularity. Our control protocol strategically engineers the sensor's time evolution so the generators for the two parameters become orthogonal. It not only removes the singularity but also restores the optimal scaling of precision with interrogation time for both parameters simultaneously. We experimentally validate this protocol using a nitrogen-vacancy center in diamond at room temperature, demonstrating the concurrent achievement of the optimal scaling for both parameters under realistic conditions.
\end{abstract}

\maketitle

\section{\label{sec:level1}Introduction}

Quantum sensors harness quantum‐mechanical effects to achieve highly sensitive measurements~\cite{Giovannetti2011,RevModPhys.89.035002}. A major frontier in this field is multiparameter estimation, where the goal is to measure several physical quantities simultaneously, each at the highest possible sensitivity. However, reaching the fundamental quantum limit for a single parameter is non-trivial; achieving this for multiple parameters at once is vastly more challenging~\cite{PhysRevA.69.022303,Imai_2007,PhysRevLett.111.070403,PhysRevLett.116.030801,PhysRevLett.125.020501,PhysRevA.94.052108,Liu_2020,Demkowicz-Dobrzański_2020,vasilyev2024optimalmultiparametermetrologyquantum,hu2024controlincompatibilitymultiparameterquantum,PhysRevLett.117.160801,Hou2016,PhysRevLett.112.103604,Taylor2013,PhysRevLett.124.060502,Polino:19,Roccia_2018,Ciampini2016,Zhou:15,doi:10.1126/sciadv.abd2986, PhysRevLett.126.070503,zhang2024distributedmultiparameterquantummetrology,kqfr-bbfx,hu2025optimalschemedistributedquantum,doi:10.1126/science.adt2442,gong2026robustmultiparameterestimationusing}. The optimal initial state, control sequence, and measurement basis for one parameter are typically incompatible with those for another. Consequently, the ultimate bound set by the quantum Cramér-Rao bound (QCRB), which depends on the quantum Fisher information matrix (QFIM), is often unattainable \cite{Helstrom1969}.

In some critical scenarios, the challenge is not merely one of incompatibility but of fundamental impossibility. This occurs when the QFIM is singular (non-invertible), a situation that precludes any joint estimation of the parameters~\cite{Liu_2020, doi:10.1142/S0219749921400049,Mihailescu_2026}. A prime and practical example is found in AC magnetometry with a linearly polarized field, described by the Hamiltonian \(H_0 = \gamma B \cos(\omega t) \sigma_x\). In this ubiquitous sensing modality, which has applications ranging from biomedical imaging to materials science~\cite{Aslam2023, Casola2018}, the generators of information for the amplitude \(B\) and frequency \(\omega\) are commuting and parallel. This renders the QFIM singular, prohibiting simultaneous estimation of the parameters. Although substantial theoretical and experimental efforts have been devoted to achieving high-precision measurements of amplitude and frequency individually~\cite{PhysRevLett.119.180801, Pang2017,Schmitt2021}, their simultaneous achievement has remained an open problem.

\begin{figure}
\centering
\includegraphics[width=\columnwidth]{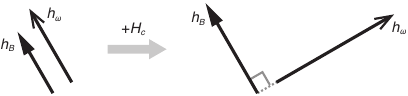}
\caption{Schematic illustration of the role of the control Hamiltonian. For the target Hamiltonian $H_\theta = \gamma B \cos(\omega t)\sigma_x$, the generators associated with $B$ and $\omega$ are commuting and parallel, resulting in a singular QFIM. Introducing the control Hamiltonian $H_c$ to optimize the time evolution makes the generators orthogonal, thereby restoring optimal scaling with respect to the sensing duration $T$.}
\label{fig:quantum_control}
\end{figure}

In this work, we propose a quantum control scheme enabling simultaneous estimation of the amplitude and frequency of a linearly polarized field. Quantum control has been previously employed in multiparameter estimation primarily to achieve optimal precision by preserving the initial direction of the generator's time evolution and maximizing its variance~\cite{PhysRevLett.117.160801, doi:10.1126/sciadv.abd2986, PhysRevLett.126.070503,kqfr-bbfx}. However, few studies have investigated its application to resolving QFIM singularities~\cite{Mihailescu_2024,PhysRevA.111.012414,doi:10.1142/S0219749921400049,Mihailescu_2026}. 
Our proposed control resolves the singularity of the QFIM by altering the directions of the generators' time evolution while maintaining the optimal scaling for each parameter. As a result, the proposed control strategy achieves sensitivities for simultaneous estimation of the amplitude and frequency of an AC field near the fundamental quantum limits in the long-time (high-frequency) regime (Fig.~\ref{fig:quantum_control}). We experimentally implement this control protocol using a nitrogen-vacancy (NV) center, a representative quantum sensing platform~\cite{Maze2008,Taylor2008,10.1063/1.3337096,Pham_2011,Wang2015,PhysRevLett.122.100501,Wang2021,PhysRevX.12.021061,PhysRevA.107.062423}. Under realistic sensing conditions, namely room temperature and low static magnetic fields, we simultaneously achieved optimal linear and quadratic sensitivity scalings for the estimation of microwave amplitude and frequency, respectively.

\section{\label{sec:level2}Theory}

Our goal is to simultaneously estimate the amplitude \(B\) and frequency \(\omega\) of an AC magnetic field encoded in the Hamiltonian \[H_0 = \gamma B \cos(\omega t) \sigma_x.\] 
As in the single-parameter case~\cite{RevModPhys.89.035002}, it is straightforward to extend our method to a general periodic function
\(
B(t) = B(a_0/2 + \sum_{n=1} A_n \cos(n\omega t + \phi_n)),
\)
as well as a specific pair $(B_m, \omega_m)$ within a multi-tone signal
\(
B(t) = \sum_m B_{m} \cos(\omega_{m} t + \phi_m).
\)

The precision of local estimation can be quantified by the quantum Cramér-Rao bound (QCRB) \cite{Helstrom1969, Liu_2020}:
\[
\textrm{Cov}(\hat{B}, \hat{\omega}) \succeq \frac{1}{M} \mathcal{F}^{-1},
\]
where \(M\) is the number of independent measurement repetitions and \(\mathcal{F}\) is the QFIM. For unitary dynamics with a pure probe state, the elements of the QFIM are given by \(\mathcal{F}_{ab} = 4\,\textrm{Cov}_{|\psi_0\rangle}(h_a(T), h_b(T))\), where the generators in the Heisenberg picture are \cite{PhysRevA.96.012117, Pang2017}:
\[
\aligned
h_a(T) &= i\,U^{\dagger}(0\to T)\,\partial_a U(0\to T)\\
&= \int_0^T U^{\dagger}(0\to t)\,\partial_a H_0(t)\,U(0\to t)\,dt.
\endaligned
\]
Given that the AC-field Hamiltonian \(H_0\) is  along \(\sigma_x\) at all times, both the derivatives, 
 \(\partial_B H_0(t) = \gamma \cos(\omega t)\sigma_x\) and \(\partial_\omega H_0(t) = -\gamma B t \sin(\omega t)\sigma_x\), and the generators \(h_B(T)\) and \(h_\omega(T)\) are also along $\sigma_x$. These commuting generators,
\(
[h_B(T), h_\omega(T)] = 0
\)
lead, for any (even entangled) probe state, to a singular QFIM of the form:
\[
\mathcal{F} \propto \begin{pmatrix} \alpha^2 & \alpha\beta \\ \alpha\beta & \beta^2 \end{pmatrix},
\]
which is non-invertible. The singularity reflects the fact that the information for \(B\) and \(\omega\) is intrinsically mixed and cannot be simultaneously extracted~\cite{Liu_2020, doi:10.1142/S0219749921400049}.

To circumvent this barrier, a control protocol needs to be designed not only to maximize the quantum Fisher information (QFI) for each parameter individually, but also to strategically reshape the geometric relationship between their generators. By making these generators nonparallel—and ideally, orthogonal—the QFIM becomes diagonal. This ensures that the precision achievable for one parameter imposes minimal constraint on the other.

To achieve this, we design a control Hamiltonian as
\[
H_c = -H(B_c,\omega_c) + \frac{\omega_c}{2} \sigma_z,
\]
where $H(B_c,\omega_c)=\gamma B_c \cos(\omega_c t) \sigma_x$ is the estimated Hamiltonian based on prior information and collected data, utilizing the estimates $B_c=\hat{B}$ and $\omega_c=\hat{\omega}$ . We specifically consider the optimal case, \( H(B_c,\omega_c)=H_0\), yielding \( U(0 \to t)=\exp{\left(-i\frac{\omega}{2}\sigma_z t\right)} \). This optimal control scenario for local estimation can be realized by adaptively refining the estimate of $B$ and $\omega$ in the control Hamiltonian~\cite{Pang2017}. Consequently, the generators become orthogonal at long times ($T \gg 2 \pi/\omega $):
\begin{align}
h_B(T)&=\gamma\int_{0}^{T} \left[ \cos^2(\omega t) \sigma_x - \sin(\omega t) \cos(\omega t) \sigma_y\right]
\,dt \nonumber\\
&\to \frac{\gamma T}{2}\sigma_x,\\
h_\omega(T)&= -\gamma B \int_{0}^{T} t\left[ \sin(\omega t) \cos(\omega t) \sigma_x - \sin^2(\omega t)  \sigma_y\right]
\,dt \nonumber\\
&\to \frac{\gamma BT^2}{4}\sigma_y.
\end{align}
The deviations from this limit decay as \((\omega T)^{-1}\)~\cite{supplementary}. Importantly, this results in orthogonal generators, $[h_{B}(T),h_{\omega}(T)]\neq0$ and $\mathrm{tr}[h_{B}(T)h_{\omega}(T)]=0$.

Using a maximally entangled Bell state as the probe, which is known to be optimal for maximizing the QFI for such non-commuting generators \cite{PhysRevLett.117.160801,kqfr-bbfx,supplementary}, the QFIM becomes diagonal:
\[\mathcal{F}=\begin{pmatrix} \gamma^2T^2 & 0 \\ 0 & \frac{1}{4}\gamma^2B^2T^4 \end{pmatrix}.
\]
Consequently, the QFIM elements for the amplitude $B$ and frequency $\omega$ exhibit the optimal quadratic ($T^2$) and quartic ($T^4$) time scalings, respectively~\cite{PhysRevLett.125.020501}.

\section{\label{sec:level3}Experiment}
\begin{figure}
\centering
\includegraphics[width=0.65\columnwidth]{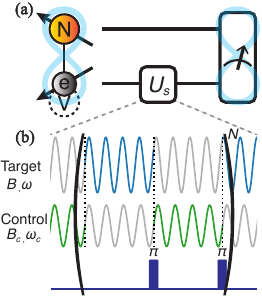}
\caption{(a) Schematic diagram of the experiment using a single NV center. The electronic spin serves as the sensor qubit, while the nitrogen nuclear spin is utilized as an ancilla qubit. The experimental sequence consists of three steps: Bell-state preparation, sensing, and Bell-state measurement. (b) Schematic diagram of the sensing sequence. The target magnetic field (blue) and control magnetic field (green) alternately interact with the sensor. Dynamical decoupling is implemented by applying $\pi$ pulses to cancel the Ising-type interaction between the sensor and ancilla qubits.}
\label{fig:experiment}
\end{figure}

\subsection{Methods}
We employ the electronic-nuclear spin system of the $^{14}$NV center~\cite{Hirose2016,Hernández-Gómez2024,kqfr-bbfx}, defining the sensor and ancilla qubits within the subspace spanned by the four levels $\{|m_S, m_I\rangle = |0,+1\rangle, |0,0\rangle, |-1,+1\rangle, |-1,0\rangle\}$. We follow the Bell-state preparation and Bell-basis readout protocol described in Ref.~\cite{kqfr-bbfx} (Fig.~\ref{fig:experiment}(a)). 
Starting from the optically polarized state $\ket{0, +1}$, an RF $\pi/2$ pulse on the nuclear spin
followed by a selective MW $\pi$ pulse on the electronic spin prepares the Bell state
\(
\ket{\Phi_{+}} = \frac{1}{\sqrt{2}}\!\left(\ket{0,0} + \ket{-1,+1}\right).
\)
After the sensing stage, the entangling operations are reversed, and the desired populations are extracted via sequential electron-spin fluorescence readout. This process is facilitated by temporarily storing population in the third level of the nitrogen nuclear spin, outside the qubit subspace. The method, operable at room temperature and under moderate static bias fields, enables averaged readout of the Bell basis without requiring single-shot readout capabilities.

 In the total sensing sequence $U_s$, we set the control frequency \(\omega_c = D - \gamma_e B^0_z-A/2\), where $D = (2\pi)\times 2.87\,\mathrm{GHz}$ is the zero-field splitting, $A = -(2\pi)\times2.16$ MHz is the hyperfine coupling constant, and $\gamma_{e}= (2\pi)\times 2.8\,\mathrm{MHz/G}$ is the gyromagnetic ratio of the electronic spin. A static magnetic field $B^0_z= 357\,$G is applied along the NV axis, resulting in the total Hamiltonian
\(
H = H_0 + H_c + H_{\mathrm{int}},
\)
where $H_\textrm{int}=A (\sigma_z^e + \sigma_z^n - \sigma_z^e \sigma_z^n)/4$ describes the hyperfine interaction between the electronic and nuclear spins.
The main challenge in this experimental implementation is the always-on interaction between the sensor and the ancilla, which we address by employing a discrete control scheme with dynamical decoupling (DD)~\cite{PhysRevLett.126.070503,kqfr-bbfx}; as illustrated in Fig.~\ref{fig:experiment}(b), alternating the interactions with the target and control magnetic fields while inserting \(\pi\) pulses cancels the unwanted sensor-ancilla interaction (see Supplemental Section~\ref{app:e}). One such cycle constitutes an interrogation block of duration \(\tau\), which we repeat \(N\) times for a total interrogation time \(T = N \tau\).

Sensor precision is evaluated via the parameter covariance matrix, given by  
\(
\mathrm{Cov}(\hat\theta)=J^{-1}\,\mathrm{Cov}(\hat p)\,(J^{-1})^{\!\top}, 
J_{ij}\equiv\frac{\partial p_i}{\partial\theta_j},
\)
where the Jacobian matrix \(J\) relates the measured signals
\(p\) to the target parameters \(\theta\).
Since our readout is photon-shot-noise limited, we approximate  
\(\mathrm{Cov}(\hat p)\simeq\sigma^{2}\mathbf I\). To avoid a flat signal at the optimal point, we rotate the measurement basis by  
\(
U_{\mathrm r}=e^{-i\pi(\sigma_x^{e}+\sigma_y^{e}+\sigma_z^{e})/3\sqrt3},
\)
which ensures an equal outcome distribution \(p_i=1/4\) and maximizes the sensitivity by increasing the magnitude of the derivatives \(|\partial p_i/\partial\theta_j|\)~\cite{kqfr-bbfx}.

\subsection{Results}
\begin{figure}
\centering
\includegraphics[width=\columnwidth]{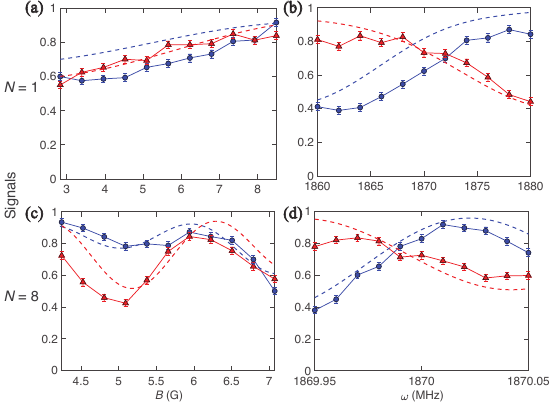}
\caption{Normalized measured signals, corresponding to $1-p_i$ ($i=1,2$; $i=1$ in blue and $i=2$ in red), as functions of the amplitude $B$ (a,c) and frequency $\omega$ (b,d) of the target microwave field for two different numbers of repetitions, $N=1$ (a,b) and $N=8$ (c,d). Here, the control field parameters were set as follows: amplitude \(B_c = 5.65\ \text{G}\) and frequency \(\omega_c = (2\pi)\times1870\ \text{MHz}\). 
The dashed lines represent the model that accounts for SPAM errors due to laser irradiation. Error bars represent the
standard deviation of the signal. Each displayed point is a result of $n=3\times10^6$ averages of sequences.}
\label{fig:sweep}
\end{figure}

\begin{figure}
\centering
\includegraphics[width=\columnwidth]{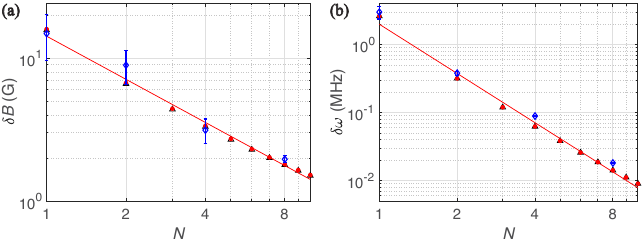}
\caption{The minimum measurable amplitude (a) and frequency (b) of the target microwave field as functions of the number of repetitions $N$. The uncertainties in the estimates of the target parameters are plotted for both the simulation (red) and the experiment (blue) at the point of maximum sensitivity, i.e., around the parameters of the control field shown in Fig.~\ref{fig:sweep}. The black triangle points represent the uncertainties calculated from the three independent populations of the Bell basis. Error bars on the experimental data indicate the standard error of the uncertainty, obtained via error propagation from the standard error of the fitted slopes. A fit to the simulation (red line) shows optimal linear and quadratic
sensitivity scalings for amplitude and frequency, respectively; see Supplemental Material for details.}
\label{fig:sensitivity}
\end{figure}

We first characterize the sensor response by varying each target parameter individually while maintaining the other at the control-field setting, as shown in Fig.~\ref{fig:sweep}. Although the readout scheme we have developed in Ref.~\cite{kqfr-bbfx} can obtain all three independent populations of the Bell basis, numerical simulations demonstrate that sensitivities derived from all three independent signals are nearly identical to those from the first two signals (Fig.~\ref{fig:sensitivity}). Thus, we restrict our analysis to two signals, simplifying experimental implementation.
For a single interrogation block ($N=1$), the sensor signals exhibit smooth variations with respect to the target parameters (Figs.~\ref{fig:sweep}(a,b)).
Increasing the number of repetitions to $N=8$ significantly enhances these variations, resulting in steeper gradients around the optimal point (Figs.~\ref{fig:sweep}(c,d)), directly translating into optimal sensitivity scaling with respect to the interrogation time $T$.

Linear fits to the data in Fig.~\ref{fig:sweep} provide the elements of
the experimentally-determined Jacobian matrix \(J_{ij}=\partial p_i/\partial\theta_j\),
from which we obtain the parameter covariance matrix \(\mathrm{Cov}(\hat\theta)\).
Figure~\ref{fig:sensitivity} shows the uncertainties
\((\delta B,\delta\omega)\) as functions of
the repetition number~\(N\).
Both experimental (blue) and simulated (red)
results confirm the expected linear scaling \(\delta B \propto N^{-1}\;(\propto T^{-1})\) and quadratic scaling \(\delta\omega \propto N^{-2}\;(\propto T^{-2})\), respectively, confirming that our control scheme successfully achieves optimal scaling despite experimental imperfections inherent to room-temperature NV center implementations.

\section{\label{sec:level5}Conclusion}

In summary, we have proposed and experimentally demonstrated a quantum control scheme that successfully resolves the singularity in the QFIM, enabling simultaneous estimation of the amplitude and frequency of linearly polarized AC fields. Unlike previous quantum control approaches focusing solely on maximizing generator's variance while preserving its direction, our method strategically alters the direction of the generator's time-evolution. This approach effectively removes the fundamental limitation arising from commuting and parallel generators, recovering optimal sensitivity scaling near the quantum limits. Experimental validation was performed using an NV center at room temperature under realistic sensing conditions, demonstrating optimal linear sensitivity scaling for amplitude estimation and quadratic scaling for frequency estimation simultaneously. Future directions include extending this strategy to more complex multiparameter quantum sensing scenarios and developing more general methods for constructing quantum control protocols to systematically address QFIM singularities.

\begin{acknowledgments}
We thank Yuichiro Matsuzaki for insightful discussions.
This work was partially supported by the National Science Foundation under
Grant Nos.~PHY-1734011, MPS-2328774 and PHY-1734011 (the MIT-Harvard Center for Ultracold Atoms), by the National Research Foundation of Korea (NRF) via the Cooperative Research on Quantum Technology (2022M3K4A1094777),
by Quantum Science and Technology-National Science and Technology Major Project (2023ZD0300600), and by the Research Grants Council of Hong Kong (14309223, 14309624, 14309022), the Guangdong Provincial Quantum Science Strategic
Initiative (GDZX2303007), 1+1+1 CUHK-CUHK(SZ)-GDST Joint Collaboration Fund (Grant No. GRDP2025-022). We acknowledge support from the ARO(W911NF-23-1-0077), ARO MURI (W911NF-21-1-0325), AFOSR MURI (FA9550-21-1-0209, FA9550-23-1-0338), DARPA (HR0011-24-9-0359, HR0011-24-9-0361), NSF (ERC-1941583, OMA-2137642, OSI-2326767, CCF-2312755, OSI-2426975), Packard Foundation (2020-71479), the Marshall and Arlene Bennett Family Research Program, quantised 2.0 and the Q-NEXT Center. This material is based upon work supported by the U.S. Department of Energy, Office of Science, National Quantum Information Science Research Centers and Advanced Scientific Computing Research (ASCR) program under contract number DE-AC02-06CH11357 as part of the InterQnet quantum networking project. T.~Isogawa acknowledges support from the Keio University Global Fellowship and a Mathworks fellowship.
\end{acknowledgments}

\bibliography{multiparam_AC}

\newpage
\clearpage
\onecolumngrid
\vspace{2cm} 

\begin{center}
    {\Large \textbf{Supplemental Material: Approaching the Limit in Multiparameter AC Magnetometry with Quantum Control}}\\[0.5cm]

\end{center}
\setcounter{section}{0}
\setcounter{equation}{0}
\setcounter{figure}{0}
\section{\label{app:a}Convergence in the long-time limit}

The elements of the quantum Fisher information matrix (QFIM) for the pure state $\ket{\psi_0}$, denoted as $\mathcal{F}_{ab}$, are given by the covariance of the generators:
\begin{equation}\label{eq:QFIM_eq}
 \mathcal{F}_{ab} =4\,\mathrm{Cov}_{\ket{\psi_0}}\!\bigl(h_{\theta_a}(T),h_{\theta_b}(T)\bigr).
\end{equation}
Here, the covariance is $\mathrm{Cov}_{\ket{\psi_0}}\!\bigl(h_{\theta_a}(T),h_{\theta_b}(T)\bigr)
 =\frac{1}{2}\langle \bigl\{h_{\theta_a}(T),h_{\theta_b}(T)\bigr\}\rangle-\langle h_{\theta_a}\rangle\langle h_{\theta_b}\rangle$ and 
the generators in the Heisenberg picture are obtained as
\begin{equation}
h_\theta(T) = \int_0^T U_\theta^\dagger(t)\, \partial_\theta H_\theta(t)\, U_\theta(t)\, dt.
\end{equation}
Under the optimal control setting, the total Hamiltonian simplifies to
\begin{equation}
H_{\text{tot}}(t) = H(t) + H_c(t) = \frac{\omega_c}{2}\sigma_z,
\end{equation}
with the corresponding time-evolution operator
\begin{equation}
U(0 \rightarrow t) = \exp\left(-\frac{i \omega_c t}{2} \sigma_z\right).
\end{equation}
This unitary dynamics yields the analytical forms of the generators:
\begin{align}
h_B(T) &= \frac{\gamma}{2} \left(T + \frac{\sin(2 \omega T)}{2 \omega}\right) \sigma_x - \frac{\gamma}{2} \left(\frac{1 - \cos(2 \omega T)}{2 \omega}\right) \sigma_y, \\
h_{\omega}(T) &= -\frac{\gamma B}{2} \left( \frac{-T \cos(2 \omega T)}{2 \omega} + \frac{\sin(2 \omega T)}{4 \omega^2} \right) \sigma_x + \frac{\gamma B}{2} \left( \frac{T^2}{2} - \frac{T \sin(2 \omega T)}{2 \omega} - \frac{\cos(2 \omega T) - 1}{4 \omega^2} \right) \sigma_y.
\end{align}
Substituting these into Eq.~\eqref{eq:QFIM_eq} with the optimal probe state $\ket{\psi_0} = \tfrac{1}{\sqrt{2}}(\ket{00}+\ket{11})$ (see Supplemental Section~\ref{app:optimal_probe_state}), the QFIM elements are obtained as
\begin{equation}
\mathcal{F}_{BB} = \frac{\gamma^2 (1 + 2 \omega^2 T^2  - \cos(2\omega T) + 2 \omega T\sin(2\omega T))}{2 \omega^2}
\end{equation}
\begin{equation}
\mathcal{F}_{B\omega} = \mathcal{F}_{\omega B} = \frac{\gamma^2 B (-1 - \omega^2 T^2 + (1 + 3 \omega^2 T^2) \cos(2\omega T))}{4 \omega^3}
\end{equation}
\begin{equation}
\mathcal{F}_{\omega \omega} = \frac{\gamma^2 B^2 (1 + 4 \omega^2 T^2  + 2 \omega^4 T^4- (1 + 2 \omega^2 T^2) (\cos(2\omega T) + 2 \omega T \sin(2\omega T)))}{8 \omega^4}
\end{equation}
The determinant of the exact $2\times 2$ QFIM admits the compact closed form
\begin{equation}
\det F
=
\frac{\gamma^{4}B^{2}T^{4}}{16\,\omega^{2}}
\Bigl(2\omega T-\sin(2\omega T)\Bigr)^{2}.
\label{eq:detF_finiteT}
\end{equation}
Since $\sin x < x$ for all $x>0$, Eq.~\eqref{eq:detF_finiteT} implies $\det F>0$ for any $T>0$ and $\omega\neq 0$ (and $B\neq 0$), i.e., the QFIM is strictly invertible under the evolution with control. 

\begin{figure}
    \centering
    \includegraphics[width=0.7\linewidth]{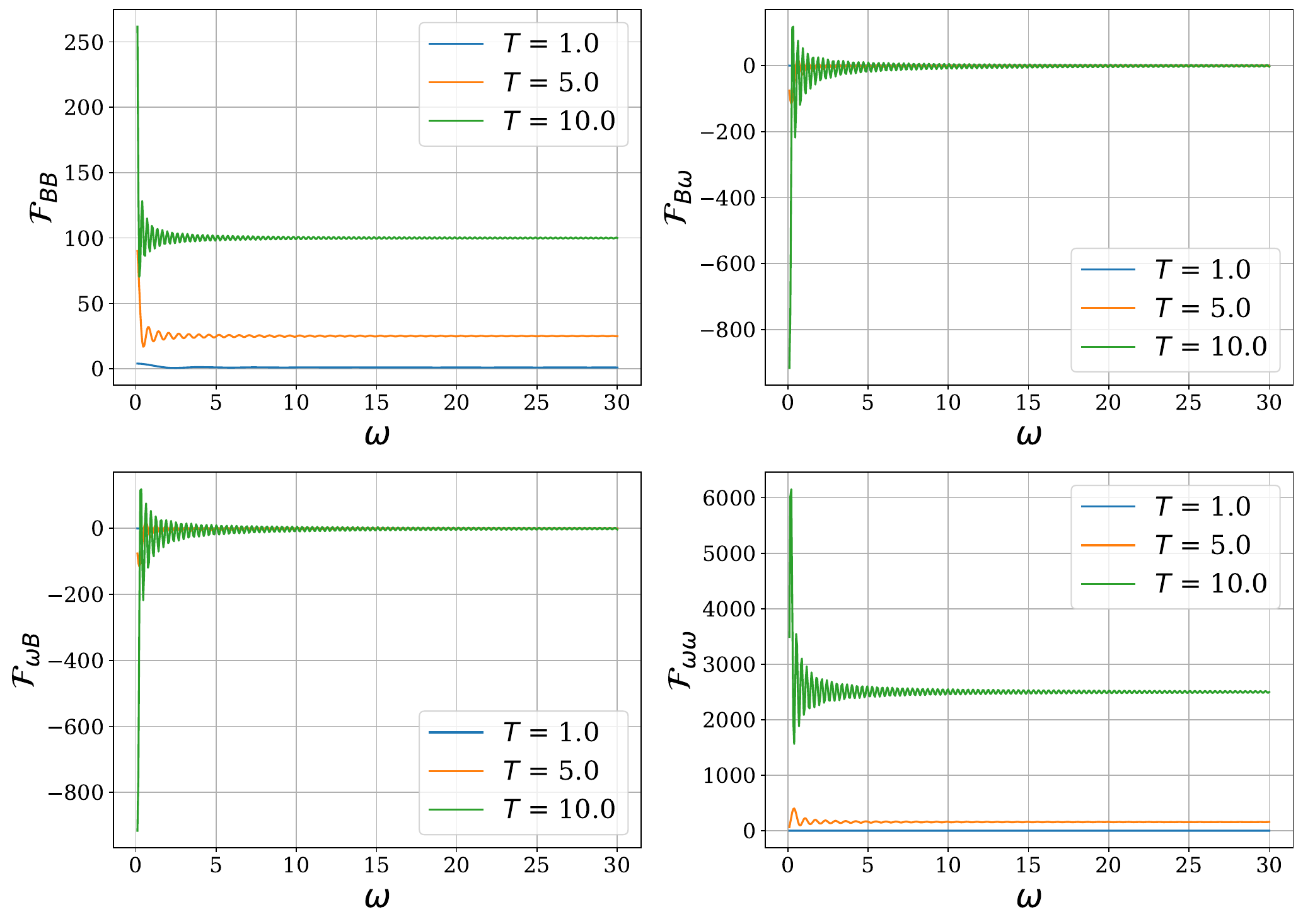}
    \caption{$\omega$ dependence of the quantum Fisher information matrix (QFIM) elements for \(T = 1\), \(5\), and \(10\). The parameter values are \(B = 1\) and \(\gamma = 1\).}
    \label{fig:QFIM_Tplot}
\end{figure}

In the limit of $\omega T \gg 2\pi$, the generators simplify to
\begin{equation}
    h_B(T)^{(\omega T\rightarrow \infty)}=\frac{\gamma T}{2}\sigma_x,\quad h_\omega(T)^{(\omega T\rightarrow \infty)}=\frac{\gamma B T^2}{4}\sigma_y
\end{equation}
and the QFIM reduces to
\begin{equation}
    \mathcal{F}_{BB}^{(\omega T\rightarrow\infty)} = 4\left(\frac{\gamma T}{2}\right)^2,\quad \mathcal{F}_{\omega\omega}^{(\omega T\rightarrow\infty)} = 4\left(\frac{\gamma BT^2}{4}\right)^2,\, \qquad \mathcal{F}_{B\omega}^{(\omega T\rightarrow\infty)}=\mathcal{F}_{\omega B}^{(\omega T\rightarrow\infty)}=0
\end{equation}
We plot the elements of the QFIM as a function of the driving frequency~\(\omega\), as shown in Fig.~\ref{fig:QFIM_Tplot} for the parameter values \(B = 1\), \(\gamma = 1\), and \(T = 1,5,10\). The plot demonstrates that each element of the QFIM approaches an asymptotic limit as the frequency increases, with the off-diagonal elements specifically vanishing in the long-time (high-frequency) regime.
Figure~\ref{fig:QFIM_Tplot} further illustrates the dependence of each element of the QFIM on the total measurement time \(T\). As the measurement time increases, the off-diagonal elements decay more rapidly with increasing frequency compared to those at shorter measurement times. 

The improvement of precision in the long-time regime can also be quantified by evaluating the relative errors of the generators and the elements of the QFIM with respect to their asymptotic limits for $\omega T\gg 2\pi$. 

\begin{figure}[h]
   \centering
   \includegraphics[width=0.8\linewidth]{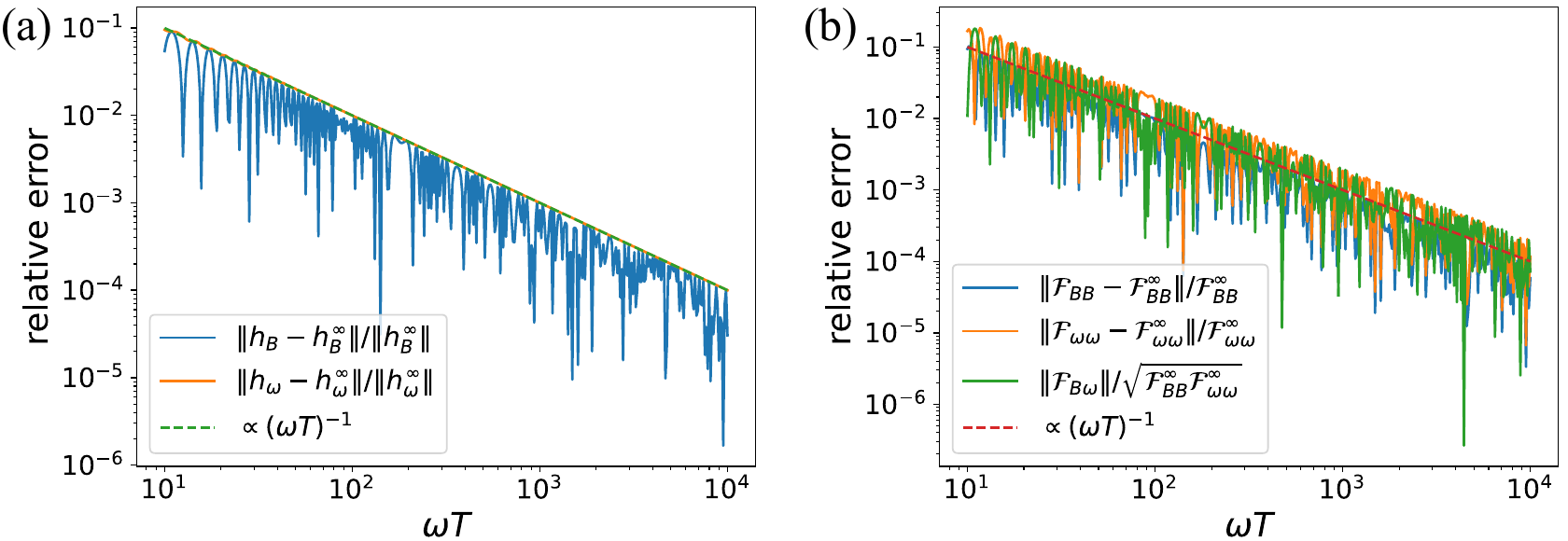}
   \caption{Relative error of the generators (a) and the QFIM elements (b) as a function of $\omega T$.}
    \label{fig:relative_error}
\end{figure}

For the generators $h_B$ and $h_\omega$, we obtain
\begin{equation}
    \frac{\|h_B-h_B^{(\omega T\rightarrow \infty)}\|}{\|h_B^{(\omega T\rightarrow \infty)}\|}\approx   \frac{\|\cos( \omega T)\|}{\omega T} ~\text{,}\quad \frac{\|h_\omega(T)-h_\omega^{(\omega T\rightarrow \infty)}\|}{\|h_\omega^{(\omega T\rightarrow \infty)}\|}\approx\frac{1}{\omega T}.
\end{equation}
This directly implies that
\begin{equation}
\frac{\|\delta h_B\|}{\|h_B^{(\omega T\rightarrow \infty)}\|}
      =\mathcal{O}\!\bigl[(\omega T)^{-1}\bigr],
\qquad
\frac{\|\delta h_\omega\|}{\|h_\omega^{(\omega T\rightarrow \infty)}\|}
      =\mathcal{O}\!\bigl[(\omega T)^{-1}\bigr].
\end{equation}
As shown in Fig.~\ref{fig:relative_error}(a), the relative error decreases with increasing $\omega T$, exhibiting a $(\omega T)^{-1}$ scaling behavior.

Next, we consider the elements of the QFIM $\mathcal{F}_{BB},~\mathcal{F}_{B\omega},~\mathcal{F}_{\omega B},\text{ and }\mathcal{F}_{\omega \omega}$. The relative errors of the matrix elements with respect to their asymptotic limits can be approximated as
\begin{equation}
    \frac{\|\delta F_{BB}\|}{\|F_{BB}^{(\omega T\rightarrow \infty)}\|}
    \approx \frac{\|\sin(2\omega T)\|}{\omega T},
\end{equation}
\begin{equation}
    \frac{\|\delta F_{\omega \omega}\|}{\|F_{\omega \omega}^{(\omega T\rightarrow \infty)}\|}
    \approx \frac{\|2\sin(2\omega T)\|}{\omega T},
\end{equation}
\begin{equation}
    \frac{\|\delta F_{B\omega}\|}{\sqrt{F_{BB}^{(\omega T\rightarrow \infty)}F_{\omega\omega}^{(\omega T\rightarrow \infty)}}}
    = \frac{\|\delta F_{\omega B}\|}{\sqrt{F_{BB}^{(\omega T\rightarrow \infty)}F_{\omega\omega}^{(\omega T\rightarrow \infty)}}}
    \approx \frac{\|3\cos(2\omega T)-1\|}{2\omega T}.
\end{equation}

We observe that all three quantities exhibit the same $(\omega T)^{-1}$ scaling, as illustrated in Fig.~\ref{fig:relative_error}(b).  
These results support the conclusion that measurement precision systematically improves with increasing $\omega T$: larger values of $\omega T$ suppress the off-diagonal elements of the QFIM, while simultaneously enhancing the diagonal elements, thereby reducing the overall estimation error.

\section{\label{app:single_parameter_benchmark}Single-parameter benchmarks and comparison with sequential strategies}

In the main text, we focus on resolving the singularity of the QFIM and on the optimal
time scalings enabled by the control protocol.
Here, we provide a benchmark based on single-parameter estimation of either
the amplitude $B$ or the frequency $\omega$ of the same target Hamiltonian
$H_0(t)=\gamma B\cos(\omega t)\sigma_x$,
and we compare our simultaneous-estimation protocol with a sequential single-parameter estimation strategy.
Throughout this section, we consider the long-time (high-frequency) regime
$\omega T\gg 2\pi$.

For unitary parameter encoding with a Hamiltonian $H_0(t)$,
the quantum Fisher information (QFI) for a single parameter $\theta$ is
$\mathcal{F}_\theta = 4\,\mathrm{Var}_{|\psi_0\rangle}\!\bigl(h_\theta(T)\bigr)$,
with the generator
$h_\theta(T)=\int_0^T U^\dagger(0\!\to\!t)\,\partial_\theta H_0(t)\,U(0\!\to\!t)\,dt$.
Maximizing over all probe states yields the QFI
$\mathcal{F}_\theta^{(\max)}$.
A convenient upper bound is given by the integrated spectral range of
$\partial_\theta H_0(t)$~\cite{PhysRevA.96.012117,Pang2017}:
\begin{equation}
\mathcal{F}_\theta^{(\max)}
\;\le\;
\left[\int_0^T\bigl(\mu_{\max}(t)-\mu_{\min}(t)\bigr)\,dt\right]^2,
\label{eq:single_gap_bound}
\end{equation}
where $\mu_{\max}(t)$ and $\mu_{\min}(t)$ are the maximum and minimum eigenvalues of
$\partial_\theta H_0(t)$.
For a qubit Hamiltonian $\partial_\theta H_0(t)=f_\theta(t)\,\sigma_x$,
this simplifies to
\begin{equation}
\mathcal{F}_\theta^{(\max)}
\;\le\;
4\left[\int_0^T |f_\theta(t)|\,dt\right]^2,
\label{eq:single_gap_bound_qubit}
\end{equation}
since $\mu_{\max}(t)-\mu_{\min}(t)=2|f_\theta(t)|$.
The bound in Eq.~\eqref{eq:single_gap_bound_qubit} is saturable by suitable control~\cite{PhysRevA.96.012117,Pang2017}.

For estimating $B$ with $\omega$ assumed known,
\begin{equation}
\partial_B H_0(t)=\gamma \cos(\omega t)\,\sigma_x,
\qquad
|f_B(t)|=\gamma|\cos(\omega t)|.
\end{equation}
Equation~\eqref{eq:single_gap_bound_qubit} gives
\begin{equation}
\mathcal{F}_B^{(\max)}
\;\le\;
4\left(\int_0^T \gamma|\cos(\omega t)|\,dt\right)^2.
\label{eq:FB_bound_exact_form}
\end{equation}
In the long-time limit $\omega T\gg 2\pi$,
\begin{equation}
\int_0^T |\cos(\omega t)|\,dt
=
\frac{2}{\pi}T+\mathcal{O}\!\left(\frac{1}{\omega}\right),
\end{equation}
and thus
\begin{equation}
\mathcal{F}_B^{(\max)}
\;\xrightarrow[\omega T\gg 2\pi]{}
\frac{16}{\pi^2}\,\gamma^2 T^2.
\label{eq:FB_single_limit}
\end{equation}
This $T^2$ scaling is attainable with appropriate control protocols and has been
discussed and demonstrated in the context of optimal AC amplitude sensing.

For estimating $\omega$ with $B$ assumed known,
\begin{equation}
\partial_\omega H_0(t)= -\gamma B\,t\sin(\omega t)\,\sigma_x,
\qquad
|f_\omega(t)|=\gamma B\,t|\sin(\omega t)|.
\end{equation}
Again from Eq.~\eqref{eq:single_gap_bound_qubit},
\begin{equation}
\mathcal{F}_\omega^{(\max)}
\;\le\;
4\left(\int_0^T \gamma B\,t|\sin(\omega t)|\,dt\right)^2.
\label{eq:Fw_bound_exact_form}
\end{equation}
In the long-time limit $\omega T\gg 2\pi$,
\begin{equation}
\int_0^T t|\sin(\omega t)|\,dt
=
\frac{1}{\pi}T^2+\mathcal{O}\!\left(\frac{T}{\omega}\right),
\end{equation}
and therefore
\begin{equation}
\mathcal{F}_\omega^{(\max)}
\;\xrightarrow[\omega T\gg 2\pi]{}
\frac{4}{\pi^2}\,\gamma^2 B^2 T^4.
\label{eq:Fw_single_limit}
\end{equation}
This corresponds to the so-called ``super-Heisenberg'' scaling for frequency estimation,
and it is attainable under suitable control strategies in the single-parameter setting~\cite{Pang2017,PhysRevLett.119.180801,Schmitt2021}.

In the main text, our control protocol yields orthogonal generators in the long-time limit,
leading to a diagonal QFIM with
\begin{equation}
\mathcal{F}_{BB}=\gamma^2T^2,
\qquad
\mathcal{F}_{\omega\omega}=\frac{1}{4}\gamma^2B^2T^4,
\qquad
\mathcal{F}_{B\omega}=0,
\end{equation}
when using an optimal maximally entangled probe state.
Comparing to the single-parameter limits in
Eqs.~\eqref{eq:FB_single_limit} and~\eqref{eq:Fw_single_limit}, we obtain
\begin{equation}
\frac{\mathcal{F}_B^{(\max)}}{\mathcal{F}_{BB}}
\;\xrightarrow[\omega T\gg 2\pi]{}
\frac{16}{\pi^2}\approx 1.62,
\qquad
\frac{\mathcal{F}_\omega^{(\max)}}{\mathcal{F}_{\omega\omega}}
\;\xrightarrow[\omega T\gg 2\pi]{}
\frac{16}{\pi^2}\approx 1.62.
\label{eq:ratio_single_multi}
\end{equation}
Equivalently, for the same number of repetitions $M$ and the same interrogation time $T$,
the standard deviations satisfy
$\delta B_{\mathrm{mp}}/\delta B_{\mathrm{sp}}=\delta\omega_{\mathrm{mp}}/\delta\omega_{\mathrm{sp}}=4/\pi\approx 1.27$,
i.e., our simultaneous-estimation protocol approaches the single-parameter benchmark
up to a constant factor while enabling joint estimation (otherwise impossible due to QFIM singularity).

A common baseline for two-parameter sensing is a sequential strategy that devotes
half of the total experimental repetitions to estimating $B$ (using a $B$-optimal protocol)
and the other half to estimating $\omega$ (using an $\omega$-optimal protocol).
Assuming a total budget of $M$ repetitions,
this yields (QCRB)
\begin{equation}
\mathrm{Var}_{\mathrm{seq}}(\hat{B}) \ge \frac{2}{M\,\mathcal{F}_B^{(\max)}},
\qquad
\mathrm{Var}_{\mathrm{seq}}(\hat{\omega}) \ge \frac{2}{M\,\mathcal{F}_\omega^{(\max)}}.
\end{equation}
In contrast, the simultaneous-estimation protocol uses all $M$ repetitions for both parameters:
\begin{equation}
\mathrm{Var}_{\mathrm{mp}}(\hat{B}) \ge \frac{1}{M\,\mathcal{F}_{BB}},
\qquad
\mathrm{Var}_{\mathrm{mp}}(\hat{\omega}) \ge \frac{1}{M\,\mathcal{F}_{\omega\omega}}.
\end{equation}
Using Eq.~\eqref{eq:ratio_single_multi}, we find
\begin{equation}
\frac{\mathrm{Var}_{\mathrm{mp}}(\hat{B})}{\mathrm{Var}_{\mathrm{seq}}(\hat{B})}
=
\frac{\mathcal{F}_B^{(\max)}}{2\,\mathcal{F}_{BB}}
\;\xrightarrow[\omega T\gg 2\pi]{}
\frac{8}{\pi^2}\approx 0.81,
\qquad
\frac{\mathrm{Var}_{\mathrm{mp}}(\hat{\omega})}{\mathrm{Var}_{\mathrm{seq}}(\hat{\omega})}
\;\xrightarrow[\omega T\gg 2\pi]{}
\frac{8}{\pi^2}\approx 0.81.
\end{equation}
Thus, despite a constant-factor reduction in the per-shot QFI compared to the
single-parameter optimum, the ability to extract information about both parameters
from every run leads to a modest advantage over sequential estimation at fixed total
interrogation-time resources.

\section{\label{app:optimal_probe_state}Optimal probe state}
Here we show how to find the optimal probe state to achieve $\mathcal{F}^{\text{max}}$ in the estimation of the magnetic field as shown in Ref.~\cite{PhysRevLett.117.160801}. Using the relation between the QFIM element $\mathcal{F}_{ij}$ and the Bures distance \(d_{\text{Bures}}\), we have \(d_{\text{Bures}}^2=\frac{1}{4}\sum_{ij}\mathcal{F}_{ij}(\rho_ {\vec{x}})dx_idx_j = 2-2f(\rho_{\vec{x}},\rho_{\vec{x}+d\vec{x}})\), where \(f(\rho_{\vec{x}}, \rho_{\vec{x}+d\vec{x}})\) denotes the fidelity between the two density matrices corresponding to infinitesimally different parameters, $\rho_{\vec{x}}$ and $\rho_{\vec{x}+d\vec{x}}$~\cite{holevo_unbiased_2011,PhysRevLett.117.160801}. From this relation, it follows that achieving optimal precision requires minimizing the fidelity \(f(\rho_{\vec{x}}, \rho_{\vec{x}+d\vec{x}})\). Here, \(\rho_{\vec{x}}\) is the evolved density matrix of the pure state $\ket{\psi_{0}}$ under the parameter $\vec{x}$, given by $U_{\vec{x}}\otimes I_A\ket{\psi_{0}}\bra{\psi_{0}}U^\dagger _{\vec{x}}\otimes I_A$. The fidelity can then be expressed as
\begin{align}
f(\rho_{\vec{x}}, \rho_{\vec{x}+d\vec{x}})&=  |\bra{\psi_{0}}U^\dagger(\vec{x},0\rightarrow T)U(\vec{x}+d\vec{x},0\rightarrow T)\otimes I_A\ket{\psi_{0}}|\\
&=|Tr(\rho_{S}U^\dagger(\vec{x},0\rightarrow T)U(\vec{x}+d\vec{x},0\rightarrow T))|,
\end{align}
where $\rho_S=\text{Tr}_A(\ket{\psi_{0}}\bra{\psi_{0}})$. The unitary operators can then be recast in a more convenient form for analysis, $U^\dagger(\vec{x},0\rightarrow T)U(\vec{x}+d\vec{x},0\rightarrow T)=e^{ia(\vec{x},d\vec{x})(\vec{k}(\vec{x},d\vec{x})\cdot\vec{\sigma})}$, where $\vec{k}(\vec{x},d\vec{x})$ is a unit vector that acts on the Pauli vector. This operator can be decomposed into a diagonal matrix $D$, where the eigenvalues are $e^{+ia(\vec{x},d\vec{x})}, \,e^{-ia(\vec{x},d\vec{x})}$. From this fact, we can rewrite $e^{ia(\vec{x},d\vec{x})(\vec{k}(\vec{x},d\vec{x})\cdot\vec{\sigma})}$ as the transformational unitary operator which depends on the unit vector, $\vec{k}$ with the diagonal matrix in the middle as $\Tilde{U}(\vec{k}(\vec{x},d\vec{x}))\,D\,\Tilde{U}^\dagger(\vec{k}(\vec{x},d\vec{x}))$. Then, we consider the fidelity between $\rho_{\vec{x}}$ and $\rho_{\vec{x}+d\vec{x}}$ by defining $\Tilde{\rho}=\Tilde{U}^\dagger(\vec{k}(\vec{x},d\vec{x}))\rho_S \Tilde{U}(\vec{k}(\vec{x},d\vec{x}))$ 
\begin{align}
f(\rho_{\vec{x}}, \rho_{\vec{x}+d\vec{x}})
&=|\mathrm{Tr}\left(\Tilde{\rho}\begin{pmatrix}
e^{ia} & 0 \\
0 & e^{-ia}
\end{pmatrix}\right)|\\
&=|\Tilde{\rho}_{11}e^{ia}+\Tilde{\rho}_{22}e^{-ia}|\\
&=\sqrt{\cos^2(a)+(\Tilde{\rho}_{11}-\Tilde{\rho}_{22})^2\sin^2(a)}.
\end{align}
The fidelity reaches its minimum when \(\Tilde{\rho}_{11} = \Tilde{\rho}_{22}\), i.e., when the subsystem is maximally mixed. In a two-level system, such a reduced state typically arises when the full system is in a maximally entangled pure state, since tracing out one qubit of a maximally entangled pair yields a maximally mixed state.

\section{\label{app:d}Optimal measurement}
The optimal measurement can be found from the saturation condition of the uncertainty principle, which dictates the product of uncertainties between generators and observables in the Heisenberg picture. Specifically, the condition can be expressed as
\begin{equation}
    (O_\theta - \langle O_\theta \rangle)\ket{\psi_0} = i\lambda \,(h_\theta - \langle h_\theta \rangle)\ket{\psi_0}, \quad \lambda \in \mathbb{R}.
\end{equation}
To satisfy this relation, one must choose an observable $O_\theta$ that corresponds to the generator $h_\theta$. For the case of a linearly polarized AC field in the long-time limit, the relevant generators are given by
\[
    h_B = \frac{\gamma T}{2}\sigma_x, 
    \qquad
    h_\omega = \frac{\gamma B T^2}{4}\sigma_y .
\]
These lead to the following relations:
\begin{align}
    (h_B - \langle h_B \rangle)\ket{\psi_0} 
        &= \frac{\gamma T}{2}\,\sigma_x \otimes I \ket{\psi_0} 
        = \frac{\gamma T}{2}\,\frac{\ket{01}+\ket{10}}{\sqrt{2}}, \\
    (h_\omega - \langle h_\omega \rangle)\ket{\psi_0} 
        &= \frac{\gamma B T^2}{4}\,\sigma_y \otimes I \ket{\psi_0} 
        = -i\frac{\gamma B T^2}{4}\,\frac{\ket{01}-\ket{10}}{\sqrt{2}}.
\end{align}
Here, we have used the assumption that the expectation values of the two generators vanish for the initial state $\ket{\psi_0} = \frac{1}{\sqrt{2}}(\ket{00}+\ket{11})$.

In this case, the corresponding observables can be chosen as
\[
    O_B = \sigma_z \otimes \sigma_y,
    \qquad
    O_\omega = \sigma_x \otimes \sigma_z ,
\]
which yield
\begin{align}
    (O_B - \langle O_B \rangle)\ket{\psi_0} 
        &= i\,\frac{\ket{01}+\ket{10}}{\sqrt{2}}, \\
    (O_\omega - \langle O_\omega \rangle)\ket{\psi_0} 
        &= \frac{\ket{01}-\ket{10}}{\sqrt{2}}.
\end{align}

From these calculations, we see that the chosen observables satisfy the saturation condition of the uncertainty principle, with
\[
    \lambda_B = -\frac{\gamma T}{2}, 
    \qquad
    \lambda_\omega = -\frac{\gamma B T^2}{4}.
\]
Furthermore, since the associated observables commute, they can, in principle, be measured simultaneously.

\section{\label{app:e}Dynamics of the NV two-qubit system}
\begin{figure}
\centering
\includegraphics[width=0.9\columnwidth]{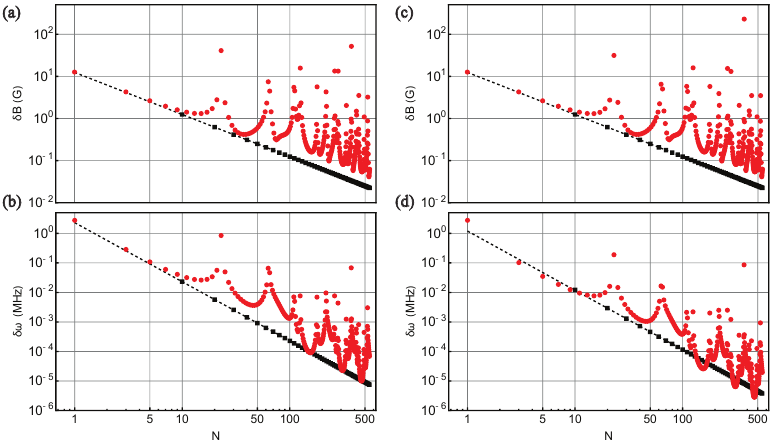}
\caption{The minimum measurable amplitude (a,c) and frequency (b,d) of the target microwave field as functions of number of repetitions $N$. The parameters used here are the same as those used in Figs.~3 and 4 of the main text. In panels (a) and (b), the time evolution of $H_{\theta} + H_{\text{int}}$ over the interval $kt \to (k+1)t$ is used, whereas in panels (c) and (d), that over the interval $2kt \to (2k+1)t$ is used ($k = 0, \ldots, N$). The black points represent simulations for ideal $\pi$ pulses, while the red points show simulated results obtained using the realistic $\pi$ pulses employed in our experiment. The black dashed lines indicate linear fits, with the scaling exponents being $1.00$ (a), $2.00$ (b), $1.00$ (c), and $2.00$ (d), respectively. }
\label{fig:exp_scaling}
\end{figure}

The Hamiltonian of the electronic-nuclear spin system of the $^{14}$NV center is given by
\[
\widetilde H_{SI} = D S_z^2 + \vec{B}(t)\cdot(\gamma_e \vec{S}+ \gamma_n \vec{I}) + Q I_z^2  + A S_z I_z,
\]
where $D = (2\pi)\times 2.87\,\mathrm{GHz}$ is the zero-field splitting, $Q = -(2\pi)\times 4.95$~MHz~is the nuclear quadrupole moment, $\gamma_{e}\approx (2\pi)\times 2.8\,\mathrm{MHz/G}$, $\gamma_{n}\approx-(2\pi)\times 0.31\,\mathrm{kHz/G}$ denote the gyromagnetic ratios of the electronic and nuclear spins, $A = -(2\pi)\times2.16$ MHz is the hyperfine coupling constant, $\vec S$ and $\vec I$ are the 
spin-1 operators of the electronic and nuclear spins, respectively. The total magnetic field $\vec B(t)$ includes a static component $B^0_z= 357\,$G  along the NV axis, a target field $B(t)=B\cos (\omega t+\phi)\,$, and a control field $B_c(t)=B_c\cos (\omega_c t+\phi_c)\,$.
We define two qubits within the subspace spanned by the four levels $\{|m_S, m_I\rangle = |0,+1\rangle, |0,0\rangle, |-1,+1\rangle, |-1,0\rangle\}$. Expressing the spin-1 operators in terms of effective spin-1/2 operators yields the two-qubit Hamiltonian,
\[
H_{SI}=-\frac{(D-\gamma_e B^0_z)}{2}\,\sigma_z^e-\frac{(Q+\gamma_nB^0_z)}{2}\,\sigma_z^n+\frac{A}{4} (\sigma_z^e + \sigma_z^n - \sigma_z^e \sigma_z^n)+\frac{\gamma_e B}{\sqrt{2}}\cos (\omega t+\phi)\,\sigma_x^e+\frac{\gamma_e B_c}{\sqrt{2}}\cos (\omega_c t+\phi_c)\,\sigma_x^e.
\]
Setting \(\omega_c=D-\gamma_e B^0_z-A/2\) and \(\gamma=\gamma_e/\sqrt{2}\), and moving to the rotating frame defined by $(Q+\gamma_nB^0_z)\sigma_z^n/2$, we obtain
\[
H=\gamma B\cos (\omega t+\phi)\,\sigma_x^e+\gamma B_c\cos (\omega_c t+\phi_c)\,\sigma_x^e-\frac{\omega_c}{2}\,\sigma_z^e+\frac{A}{4} (\sigma_z^e + \sigma_z^n - \sigma_z^e \sigma_z^n).
\]
To analyze the dynamics, we further move to the rotating frame defined by $\omega_c\,\sigma^e_z/2$. The Hamiltonian becomes
\[H_{\mathrm{tot}}=\underset{H_\theta}{\underline{%
  \gamma B[\cos((\omega-\omega_c)t+\phi)\sigma^e_x-\sin((\omega-\omega_c)t+\phi)\sigma^e_y]}}
\;+\;
\underset{H_c}{\underline{%
  -\gamma B_c[\cos(\phi_c)\sigma^e_x-\sin(\phi_c)\sigma^e_y]}}
\;+\;
\underset{H_{\mathrm{int}}}{\underline{%
  \tfrac{A}{4}\bigl(-\sigma_z^{e}+\sigma_z^{n}-\sigma_z^{e}\sigma_z^{n}\bigr)}}\]

Our goal is to remove the effect of $H_{\text{int}}$ and realize the dynamics given by \(H_\theta+H_c\). We employ a discrete control scheme with dynamical decoupling~\cite{PhysRevLett.126.070503,kqfr-bbfx}. The contribution proportional to $\sigma_z^{n}$ can be canceled by adjusting the phase of the second nuclear $\pi/2$ pulse (see Ref.~\cite{kqfr-bbfx}); hence we neglect $A\sigma_z^{n}/4$ below. We then address the remaining terms $\sigma_z^e$ and $\sigma_z^e\sigma_z^n$ by defining
\(
H_{\text{int}}'=A/4(-\sigma^e_z-\sigma^e_z\sigma_z^n),
\)
and applying electronic $\pi$ pulses $U_\pi=\sigma^e_x$. Conjugation by $\sigma^e_x$ reverses the sign of $\sigma^e_z$ and $\sigma^e_z\sigma_z^n$ via $\sigma_x\sigma_z\sigma_x=-\sigma_z$, yielding
\[
\sigma^e_x (H_{c}+ H_{\text{int}}')\,\sigma^e_x
= -\gamma B_c\!\left[\cos(\phi_c)\sigma^e_x+\sin(\phi_c)\sigma^e_y\right]
  - H_{\text{int}}'.
\]
Choosing $\phi_c=-\phi$ aligns the control phase with the target-field phase,
\[
\sigma^e_x (H_{c}+ H_{\text{int}}')\,\sigma^e_x
= -\gamma B_c\!\left[\cos(\phi)\sigma^e_x-\sin(\phi)\sigma^e_y\right]
  - H_{\text{int}}'.
\]
We partition the total interrogation time $T$ into small intervals of length $\tau = T/N$. By alternating the time evolutions generated by $H_{\theta} + H'_{\mathrm{int}}$ and $H_c - H'_{\mathrm{int}}$ over each interval $k \tau \rightarrow (k+1) \tau$, we can approximately reproduce the overall evolution governed by $H_{\theta} + H_c$ during the full duration $0 \rightarrow T$ (Fig.~\ref{fig:exp_scaling}(a,b)). In general, the time evolution within an interval $(k-1)\tau \rightarrow k \tau$ may explicitly depend on the index $k$ if the Hamiltonian is time-dependent. However, since we have moved into a rotating frame with a frequency matched to that of the control Hamiltonian, the control term becomes independent of $k$.
It is important to note that this control protocol doubles the total duration of the sensing sequence to $2T$. In our experiment, to reflect a more realistic sensing scenario, the target magnetic field is applied during intervals $[2k \tau, (2k+1)\tau]$, while the control magnetic field is applied in the subsequent intervals $[(2k+1)\tau, (2k+2)\tau]$ (see Fig.~2 in the main text). This setup models the practical constraint that the target field cannot be switched on and off arbitrarily. Despite this modification, the scheme exhibits the same scaling behavior, as shown in Fig.~\ref{fig:exp_scaling}(c,d).

Note that while an ideal $\pi$-pulse corresponds to applying $\sigma^e_x$, its practical implementation is challenging due to hyperfine coupling effects. These effects accumulate as the number of pulses $N$ increases, leading to oscillations in sensitivity as a function of $N$ (see Fig.~\ref{fig:exp_scaling}). Techniques such as composite pulses can mitigate these effects, enabling the realization of near-ideal $\pi$-pulses even in the presence of hyperfine coupling.

\end{document}